\documentclass[prl,twocolumn,amsmath,amssymb,amsfonts,superscriptaddress]{revtex4-2}

\usepackage{graphicx}
\usepackage{xcolor}
\usepackage{comment}

\usepackage{glossaries}
\newacronym{SCGLE}{SCGLE}{self-consistent generalized Langevin equation theory}
\newacronym{NESCGLE}{NE-SCGLE}{non-equilibrium self-consistent generalized Langevin equation theory}
\newacronym{MCT}{MCT}{mode-coupling theory}
\newacronym{SBR}{SBR}{stochastic $\beta$-relaxation theory}
\newacronym{ITT}{ITT}{integration-through transients}
\newacronym{FDT}{FDT}{fluctuation-dissipation theorem}
\DeclareMathOperator{\smol}{\Omega}

\begin{document}

\title{From sub-aging to hyper-aging in structural glasses}
\author{Luis F. Elizondo-Aguilera}
\affiliation{Institut f\"ur Materialphysik im Weltraum, 
Deutsches Zentrum f\"ur Luft-und Raumfahrt (DLR), Linder H\"ohe 
51170 K\"oln, Germany}

\author{Tommaso Rizzo}
\affiliation{Dipartimento di Fisica, Universit\`a di Roma~I ``La Sapienza'',
Piazzale A.~Moro~2, I-00185 Rome, Italy}
\affiliation{ISC-CNR, UOS Roma, Universit\`a di Roma~I ``La Sapienza'',
Piazzale A.~Moro~2, I-00185 Rome, Italy}

\author{Thomas Voigtmann}
\email{thomas.voigtmann@dlr.de}
\affiliation{Institut f\"ur Materialphysik im Weltraum, 
Deutsches Zentrum f\"ur Luft-und Raumfahrt (DLR), Linder H\"ohe 
51170 K\"oln, Germany}
\affiliation{Department of Physics, Heinrich-Heine-Universit\"at,
Universit\"atsstra{\ss}e 1, 40225 D\"usseldorf, Germany}

\date{\today}

\begin{abstract}
We demonstrate {non-equilibrium} scaling laws for the aging dynamics
in glass formers
that emerge from combining a recent application of Onsager's theory
of irreversible processes with the {equilibrium} scaling laws
of glassy dynamics.
Different scaling regimes are predicted for the evolution of the system's
structural relaxation time $\tau$ with age (waiting time $t_w$),
depending on the depth of the quench from the liquid into the glass:
\emph{simple aging} ($\tau\sim t_w$) applies for quenches close to the critical
point of \gls{MCT} and implies \emph{sub-aging} ($\tau\approx\ t_w^\delta$ with $\delta<1$)
as a broad cross-over for quenches to nearly-arrested equilibrium states;
\emph{hyper-aging} (or \emph{super-aging}, $\tau\sim t_w^{\delta'}$ with $\delta'>1$) emerges for quenches deep
into the glass.
The latter is cut off by non-mean-field fluctuations that we account for
within a recent extension of \gls{MCT}, the \gls{SBR}.
We exemplify the scaling laws by a schematic model that allows to
quantitatively fit recent simulation results
for density-quenched hard-sphere-like particles.
\end{abstract}

\maketitle
\glsresetall


The response of a viscous fluid to a sudden change in control parameters
reveals a rich phenomenology as the system adapts to this change.
If the time scale of structural relaxation in the fluid, $\tau$, is
large, a slow evolution of both static and dynamic properties of the fluid
with system age (\textit{i.e.}, the waiting time $t_w$ after the quench)
is observed. For kinetically arrested states such as glasses, this
aging dynamics implies
that the properties of the material depend on the protocol of its
fabrication, as a clear signature of the non-equilibrium process
\cite{Micoulaut.2016,Cangialosi.2013,McKenna.2017}.
Hence, the understanding of the relevant non-trivial time scales in aging is
of fundamental interest for theoretical physics
and materials science alike \cite{Cugliandolo.1993,Berthier.2011,Dyre.2018}. 

This concerns in particular empirical scaling relations that have been
observed in various experimental and simulation studies on systems that
are widely different on the microscopic scale,
ranging from, \textit{e.g.}, molecular and polymeric glasses \cite{Lunkenheimer.2005,Dyre.2010a,Hecksher.2015,riechers,Purnomo.2008,Di.2011,Wagner.2017}, 
colloidal systems
\cite{Masri.2010,Jacob.2019,Martinez.2008,Guo.2011}, 
metallic alloys \cite{Ruta.2013,Ruta.2017,Ketkaew.2020},
laponite suspensions 
\cite{Angelini.2014,Jabbari.2004}, and spin glasses \cite{Rodriguez.2003,Belletti.2008,BaityJesi.2017}.
For instance,
\emph{simple} aging ($\tau\sim t_w$) and \emph{sub}-aging ($\tau\sim t_w^\delta$ 
with $\delta<1$) are commonly found;
\emph{hyper}-aging ($\tau\sim t_w^{\delta'}$, 
with $\delta'>1$) is present as an intermediate law \cite{Masri.2010,Di.2011}
and was explicitly reported, for example, in laponite 
suspensions (with exponent $\delta'\approx1.8$) \cite{Bandyopadhyay.2004} and 
dense colloidal gels ($\delta'\approx1.37$) \cite{Bissig.2003}. 
The physical mechanisms behind these scaling laws, and in particular their
relation to the microscopic details of the fluid and/or the protocol
of the control-parameter quench, so far remained unresolved.

Here we establish \emph{non-equilibrium} scaling laws of aging and show how
they emerge from the \emph{equilibrium} scaling laws for structural relaxation
near the glass transition, specifically near the critical point of
\gls{MCT}.
In principle this directly relates non-equilibrium aging exponents
to the equilibrium structure of a glass-forming material.
The predictions arise from combining two recent
theoretical approaches to describe the dynamics of glass-forming fluids:
the \gls{NESCGLE} provides a starting point linking the
waiting-time evolution of static properties to the relaxation dynamics
of the system, while a recent extension of \gls{MCT}, the
\gls{SBR} provides scaling laws for this relaxation dynamics that
also include the effect of non-mean-field fluctuations in the
ideal glass of \gls{MCT}.
We demonstrate the scaling laws by a quantitative
comparison to recent computer-simulation results for the evolution
following density-quenches in hard-sphere-like systems
\cite{Perez.2011,Mendoza.2017}
that elucidate three different scaling regimes predicted by the theory.

\Gls{MCT} is a microscopic theory \cite{Goetze.2009,Janssen.2018}
that very successfully
describes the liquid-state dynamics close to the glass transition.
In its original form it is restricted to the equilibrium ensemble,
although recent extensions allow to treat nonlinear response to various external fields \cite{Fuchs.2002c,Fuchs.2009,Gazuz.2009,Kranz.2018,Liluashvili.2017}.
Its application to aging dynamics
has been proposed 20 years ago by Latz \cite{Latz.2000,Latz.2002},
but the complexity of that theory has so far only allowed to obtain
some results linked to the seminal work by Cugliandolo \textit{et~al.}\
\cite{Cugliandolo.1993,Cugliandolo.1994,*Cugliandolo.1995b,*Cugliandolo.1995} on the $p$-spin model \cite{Kim.2001}.
The complexity stems from the fact that in absence of
the equilibrium \gls{FDT}, correlation and response functions are not
straightforwardly connected,
and are described by coupled integral equations that
are not readily evaluated.

To cut this Gordian knot,
the \gls{NESCGLE} \cite{RamirezGonzalez.2010,RamirezGonzalez.2010b,SanchezDiaz.2013}
invokes an assumption of
``local stationarity'' for the relaxation
process, reducing the complexity of the full problem considerably.
Essentially, it partially decouples the evolution of the correlation functions
from that of the underlying static response functions.
The resulting theory tests favorably against both simulation
\cite{Perez.2011,Mendoza.2017,ElizondoAguilera.2021}
and experimental data \cite{OlaisGovea.2015,OlaisGovea.2018,OlaisGovea.2019}.

\Gls{NESCGLE} in fact refers to two separate ingredients:
an evolution equation for the static obervables, and an underlying
kinetic theory for the mobility of rearrangements, the \acrshort{SCGLE}
\cite{YeomansReyna.2007}.
The latter is, for the present purposes,
structurally identical to \gls{MCT}.
In particular, it provides the same asymptotic scaling laws for the
equilibrium structural relaxation \cite{ElizondoAguilera.2019}.
We will use those well-established scaling laws to describe the asymptotic
waiting-time dependence after a quench.

%

The non-equilibrium extension of the \acrshort{SCGLE} is usually
derived by referencing Onsager's laws of linear irreversible
thermodynamics and the corresponding stochastic theory of thermal
fluctuations
(see Refs.~\cite{MedinaNoyola.1987,MedinaNoyola.1987b}).
Under certain assumptions, it leads to an innocuous looking relaxation
equation for the waiting-time evolution of the non-equilibrium static
structure factor $S(k;t_w$).
We demonstrate that this equation can also be rationalized in a spirit closer to
\gls{MCT} employing the \gls{ITT} formalism \cite{Fuchs.2002c}:
writing the evolution equation of
the non-equilibrium distribution function $p(t)$ of a system as
$\partial_tp(t)=\smol(t)p(t)$, with some linear differential operator
$\smol(t)$, a formal solution is
$p(t)-p(t_w)=\int_{t_w}^tdt'\exp_+[\int_{t'}^t\smol(\tau)\,d\tau]\mathcal P_2\smol(t')p(t_w)$
where $\mathcal P_2=1$ is the identity operator. Note that
$S(k;t_w)=(\varrho_{-\vec k}\varrho_{\vec k},p(t_w))$ where
$\varrho_{\vec k}$ are the microscopic number-density fluctuations and
$(f,g)$ is the usual $L_2$ scalar product in Hilbert space.
For a sudden quench,
$\smol(t)=\smol_i$ for $t<0$ and $\smol(t)=\smol_f$ for $t>0$, we
can make use of the relation $\smol(t')p(t_w)=\partial_{t_w}p(t_w)$
for all $t'\ge t_w>0$, which avoids the need to formulate the effect of
the quench in the time-evolution operator explicitly.
Projecting onto density-pair modes as the relevant variables,
$\mathcal P_2=\varrho_{\vec k}\varrho_{-\vec k}p(t_w))
\mathcal N^{-1}(\varrho_{-\vec k}\varrho_{\vec k}$ (suitably normalized),
and neglecting memory effects,
we obtain $S(k;t)-S(k;t_w)
\approx\int_{t_w}^tdt'\,C_4(k;t,t')\,\partial_{t_w}S(k;t_w)$ with some
four-point density correlation function $C_4(k;t,t')$, and thus for $t\to\infty$,
\begin{equation}\label{eq:sk}
  \frac{\partial S(k;t_w)}{\partial t_w}
  =-\mu(k;t_w)\left(S(k;t_w)-S_f(k)\right)\,,
\end{equation}
where $\mu(k;t_w)$ is a mobility factor that is slaved to the
structural relaxation dynamics \cite{ZepedaLopez.2021,Lunkenheimer.2005}.
The initial state before the quench is $S(k;0)=S_i(k)$, and
$S_f(k)$ characterizes the quenched-to final state.
Equation~\eqref{eq:sk} essentially is a formalized extension of the empirical
Tool-Narayanaswamy model of physical aging \cite{jcp}.

Equation~\eqref{eq:sk} already predicts universal scaling laws for the
aging dynamics to be encoded in the equilibrium dynamics:
since the glass transition is a dynamical phenomenon,
in its vicinity the static structure functions remain regular,
and we can linearize $S(k;t_w)$ for small control-parameter
distances $\varepsilon(t_w)$ to the transition.
The temporal evolution is thus asymptotically governed by the
evolution of the distance parameter along the relevant direction
in $k$-space (\gls{MCT}'s critical eigenvector \cite{Goetze.2009,Goetze.1995b}),
\begin{equation}\label{eq:epstw}
  \partial_{t_w}\varepsilon(t_w)=-\mu(\varepsilon(t_w))\left(\varepsilon(t_w)
  -\varepsilon_f\right)\,.
\end{equation}
Now enter the scaling laws for $\mu(\varepsilon)$: close to the critical
point of \gls{MCT},
$\mu(\varepsilon)\sim1/\tau(\varepsilon)\sim(-\varepsilon)^\gamma$ for liquid states
($\varepsilon<0$), and $\mu(\varepsilon)=0$ in the ideal-glass state
($\varepsilon\ge0$).
The non-trivial exponent $\gamma$ is related to the equilibrium structure of the
system at its glass transition through the \gls{MCT} exponent parameter
$\lambda$ \cite{Goetze.2009,ElizondoAguilera.2019}.
The fact that $\mu$ approaches zero, allows for non-equilibrium stationary
solutions of Eq.~\eqref{eq:epstw}, where the relaxation towards equilibrium
gets ``stuck''.

We immediately get two important scaling laws from Eq.~\eqref{eq:epstw}:
(\textit{i}) for quenches close to the glass-transition point
($|\varepsilon_f|\ll|\varepsilon_i|$),
there exists a growing window in $t_w$, where
$\partial_{t_w}\varepsilon\sim|\varepsilon|^{\gamma+1}$,
which results in $|\varepsilon|\sim t_w^{-1/\gamma}$ and, thus,
\emph{simple} or \emph{full} aging, $\tau\sim t_w$ as $t_w\to\infty$.

(\emph{ii}) for a deep quench into the ideal glass, 
$\varepsilon_f\gg|\varepsilon(t_w)|$ holds in the limit of $t_w\to\infty$, 
because the relaxation gets stuck around values close to zero. Then,
$\partial_{t_w}\varepsilon\sim|\varepsilon|^\gamma$, resulting in
the asymptotic law $\tau\sim t_w^{\gamma/(\gamma-1)}$. Since
$\gamma>1$, the exponent $\delta'=\gamma/(\gamma-1)$ is also larger than unity,
and we find \emph{hyper}-aging or \emph{super-}aging, $\tau\sim t_w^{\delta'}$ for
$t_w\to\infty$.

These scaling laws describe the idealized indefinite aging of a system
that is quenched to a state with infinite relaxation time. In reality,
the ultimate \gls{MCT}-like divergence of the relaxation time
is not observed; this one can attribute to long-wavelength
fluctuations that cause deviations from the mean-field like scenario
\cite{Rizzo:2014,Rizzo:2016,Rizzo.2020}.
It will provide a cut-off for the scaling laws, rendering them
transient rather than truly infinite-waiting-time asymptotes,
as we shall discuss below.

For quenches to liquid states close to the glass
transition, $\varepsilon_f<0$, the mobility always remains positive,
and the corresponding
long-time asymptote is then (\emph{iii})
$\tau\sim\text{const.}$ for $t_w\to\infty$.
For the typical slow evolution of the structural relaxation time,
this implies a broad cross-over where $\tau$ grows sublinearly
with $t_w$, and hence \emph{sub}-aging. Although not a rigorous
asymptote, an empirical power law, $\tau\approx t_w^\delta$ with $\delta<1$,
typically fits well in this regime \cite{Kim.2001}.

To elucidate the emergence of the three regimes -- \emph{simple}, 
\emph{sub-} and \emph{hyper}-aging -- we devise a \emph{schematic model} of 
aging. Qualitatively, the mobility is the inverse of an
integrated friction memory kernel; in the spirit of \gls{MCT} schematic
models, we assume that the slow dynamics of all such microscopic
correlation functions is governed by a single-mode (density) correlation
function $\phi(t;t_w)$,
\begin{subequations}\label{eq:sma}
\begin{equation}\label{eq:tauschem}
  \mu(t_w)=1\Big/\int_0^\infty dt\,\phi(t;t_w)\,.
\end{equation}
The latter obeys a Mori-Zwanzig type integral equation,
\begin{equation}\label{eq:mz}
  \partial_t\phi(t;t_w)+\phi(t;t_w)+\int_0^tm(t-t';t_w)\partial_{t'}\phi(t';t_w)\,dt'=0\,.
\end{equation}
In Eq.~\eqref{eq:mz} we anticipate that $t_w$ only enters parametrically
in determining the coupling coefficients of the memory kernel $m(t;t_w)$.
This encodes the assumption of local stationarity, and is in the spirit
of the \gls{ITT} framework \cite{Fuchs.2009}
that relates non-equilibrium transport coefficients to such
``transient'' correlation functions.

We complete the schematic model by the closure
\begin{equation}\label{eq:m12}
  m(t;t_w)=v_1(t_w)\phi(t;t_w)+v_2(t_w)\phi(t;t_w)^2\,,
\end{equation}
with two coupling parameters $v_1$ and $v_2$ that describe the current
$t_w$-dependent state of the system.
For fixed $t_w$, the model specified by Eqs.~\eqref{eq:mz} and \eqref{eq:m12}
is the widely studied schematic F\textsubscript{12} model of MCT.
It has a line of glass transitions $(v_1^c,v_2^c)$ where $\varepsilon=0$.
\end{subequations}

Equations \eqref{eq:sma} define our schematic model.
Together with the (mean-field) assumption $\tau(t_w)\propto1/\mu(t_w)$,
and $v_1=v_1^c$, $v_2(t_w)=v_2^c(1+\varepsilon(t_w))$ to define the distance
to the glass transition point, it allows to fit available computer-simulation
data for $\tau(t_w)$ after mapping $\varepsilon_i=\varepsilon(0)$ and $\varepsilon_f$
to the simulation's control parameters.


\begin{figure}
\includegraphics[width=\linewidth]{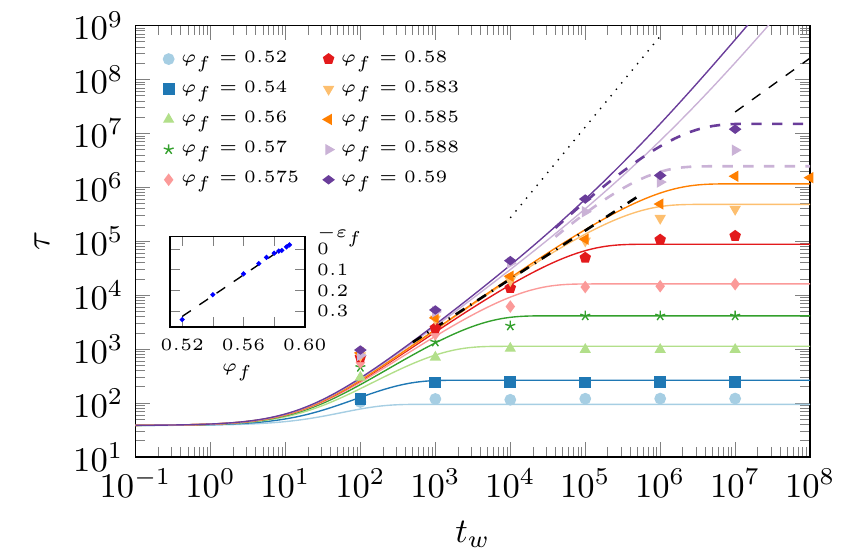}
\caption{\label{fig:tau-vs-tw}
  Structural relaxation time $\tau$ as a function of
  waiting time $t_w$ after an instantaneous quench.
  Solid lines: schematic model, quenches from $\varepsilon_i=-0.5$
  to $\varepsilon_f=-0.34$, $-0.22$, $-0.12$, $-0.07$, $-0.04$,
  $-0.02$, $-0.01$, $-0.007$, $0.01$, and $0.02$ (bottom to
  top).
  A dashed line indicates simple aging, $\tau\sim t_w$, a dotted
  line hyper-aging, $\tau\sim t_w^{\delta'}$ with $\delta'=1.684$,
  and a dash-dotted line sub-aging, $\tau\approx t_w^\delta$
  with $\delta=0.9$.
  Thick dashed lines: \acrfull{SBR} for $\varepsilon=0.01$ and $0.02$.
  Symbols: simulation results for quasi-hard spheres
  from Ref.~\protect\cite{Mendoza.2017}, quenched to various
  final packing fractions $\varphi_f$ (related to $\varepsilon_f$ as shown
  in the inset),
  translated to schematic-model units ($\tau\mapsto2\tau$, $t_w\mapsto100 t_w$).
}
\end{figure}

Results for $\tau(t_w)$ from the schematic model for quenches to
various final states close to the \gls{MCT} transition give a consistent
description of computer-simulation data for density-quenched
quasi-hard spheres (Fig.~\ref{fig:tau-vs-tw}).
For the fit, we have allowed to adjust a global time scale and the
proportionality factor between $\mu$ and $1/\tau$, and we have chosen
a transition point $(v_1^c,v_2^c)$ such that the exponent parameter of
\gls{MCT} matches a value usually found for hard-sphere like systems,
$\lambda=0.735$. This determines the exponent $\gamma=1/2a+1/2b$ with
$\Gamma(1-a)^2/\Gamma(1-2a)=\lambda=\Gamma(1+b)^2/\Gamma(1+2b)$,
and thus the exponent $\delta'\approx1.684$.

The schematic model elucidates the three aging regimes of the ideal-glass theory:
empirical sub-aging is found as a cross-over for quenches to final states in the
liquid, $\varepsilon_f<0$, while hyper-aging emerges from the model
as the asymptote for quenches to the glass, $\varepsilon_f>0$.
A growing intermediate-$t_w$ window that extends to $t_w\to\infty$ at
the critical point of \gls{MCT}, $\varepsilon_f=0$, displays simple aging.

\begin{figure}
\includegraphics[width=\linewidth]{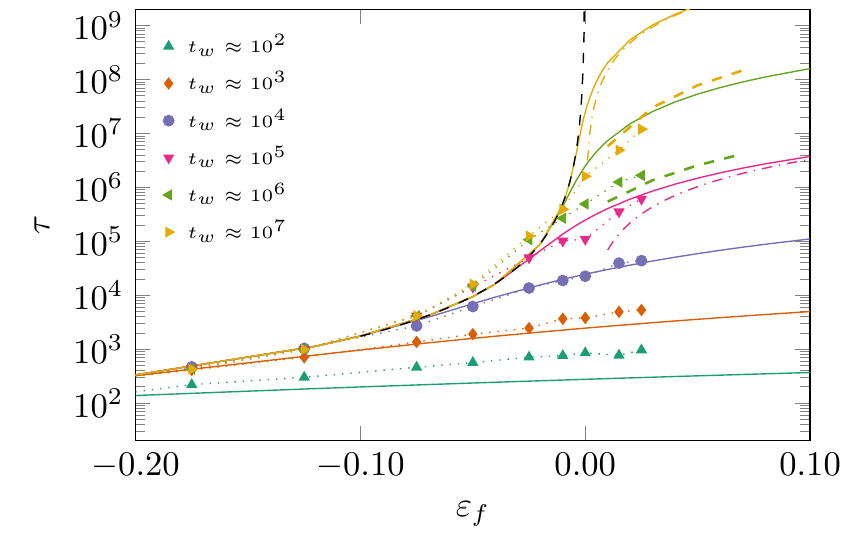}
\caption{\label{fig:tau-vs-eps}
  Structural relaxation time $\tau$ at various fixed waiting times $t_w$
  as a function of the final point of the quench $\varepsilon_f$ (solid lines and symbols:
  theory and simulation
  as in Fig.~\protect\ref{fig:tau-vs-tw}; thick dashed lines from \acrshort{SBR}).
  The thin dashed line indicates the equilibrium divergence at the ideal glass
  transition point ($\varepsilon_f=0$),
  $\tau_\text{eq}\sim|\varepsilon_f|^{-\gamma}$ with $\gamma=2.46214$,
  dash-dotted lines the non-equilibrium asymptotes
  $\tau\sim B(t_w)\varepsilon_f^{\delta'}$.
}
\end{figure}

The evolution of $\tau$ after the quench relates to the well-known problem
of determining a diverging relaxation time at fixed waiting time $t_w$
(corresponding to a typical experiment duration or probing time scale):
approaching the transition, the power-law divergence of $\tau$ as a function
of quenched-to state $\varepsilon_f$ that is predicted by the
idealized theory, is cut off at any finite $t_w$, and replaced by a cross-over
to a slower growth (Fig.~\ref{fig:tau-vs-eps}).
In our model, we obtain
$\tau\sim|\varepsilon|^{\delta'}$, with
a prefactor that diverges with increasing $t_w$ (dash-dotted lines
in Fig.~\ref{fig:tau-vs-eps}).

\begin{figure}
\includegraphics[width=.9\linewidth]{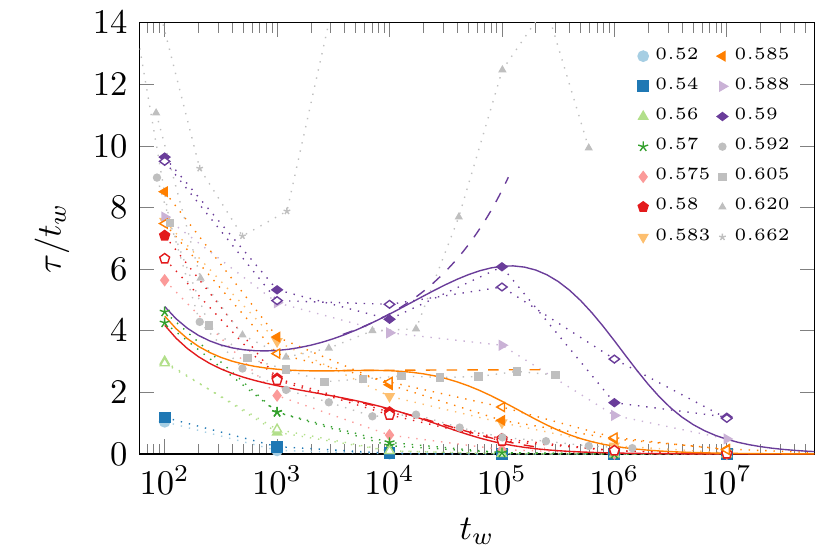}
\caption{\label{fig:tauratio-vs-tw}
  Ratio $\tau/t_w$ as a function of waiting time $t_w$.
  Simulation data of Ref.~\protect\cite{Mendoza.2017} (filled symbols;
  data divided by $50$; $\varphi_c\approx0.585$) and of Ref.~\protect\cite{Perez.2011} (open symbols);
  and on a more polydisperse system from Ref.~\protect\cite{Masri.2010} (grey;
  $\varphi_c\approx0.605$).
  Exemplary SBR results are shown for $\varepsilon_f=-0.1$, $0$, and $0.1$
  (solid lines; bottom to top). Dashed lines indicate the corresponding
  asymptotic expressions for the ideal-glass MCT.
}
\end{figure}

Deviations from the ideal theory are noted in the simulation data
for quenches to the highest final densities and at large $t_w$.
We attribute this to the avoidance of the ideal \gls{MCT} transition,
that also causes the hyper-aging regime to be interrupted.

To understand this, we turn to the \gls{SBR} \cite{Rizzo:2014,Rizzo:2016},
a recent extension of \gls{MCT} that includes fluctuations in the local
glassiness, viewing
$\sigma\sim\varepsilon$ as a dynamical fluctuating order parameter.
\Gls{SBR} predicts scaling laws that replace the divergent power law
with a cross-over between a power law on the liquid
side and exponential growth on the glassy side of the transition.
Specifically \cite{Rizzo:2015}, for the structural relaxation time
\begin{subequations}\label{eq:sbr}
\begin{equation}\label{eq:sbrtau}
  \tau\sim\left[\int_{-\infty}^0\frac{ds}{\sqrt{2\pi}\Delta\sigma}
  e^{-\frac{(s-\sigma)^2}{2\Delta\sigma^2}}|s|^{b\gamma}\right]^{-1/b}
\end{equation}
and for the mobility
\begin{equation}
  \mu\sim\int_{-\infty}^0\frac{ds}{\sqrt{2\pi}\Delta\sigma}
  e^{-\frac{(s-\sigma)^2}{2\Delta\sigma^2}}|s|^\gamma\,,
\end{equation}
\end{subequations}
where we have identified $\sigma=\varepsilon$.
Here, $\Delta\sigma$ is a material parameter that quantifies the
strength of long-wavelength order-parameter fluctuations.
Using Eqs.~\eqref{eq:sbr} to evaluate $\mu$ in Eq.~\eqref{eq:epstw} and
to calculate $\tau$, we obtain an improved asymptotic description
of the $\tau$-vs-$t_w$ curves (colored dashed lines in Fig.~\ref{fig:tau-vs-tw})
that account for the cross-over from hyper-aging to a constant $\tau$
as the system finally equilibrates even in the ideal-\gls{MCT} glass.

Interestingly, the hyper-aging law predicted by the ideal theory
still survives as a transient. In the simulation data, this
is best seen as a non-monotonic
variation of the ratio $\tau/t_w$ as a function of $t_w$
that is present
for all quenches to $\varphi_f>\varphi_c$
(Fig.~\ref{fig:tauratio-vs-tw}).
This transient hyper-aging signature fits well the
corresponding \gls{SBR} prediction (solid lines in Fig.~\ref{fig:tauratio-vs-tw}).


In conclusion, we present scaling laws for the evolution of the structural
relaxation time $\tau$ as a function of system age $t_w$ after the quench
of a glass-forming fluid to states close to the ideal glass-transition
point of \gls{MCT}.
Based on the \gls{NESCGLE} to describe the evolution of static quantities
after such quenches, the scaling laws delineate regimes of
simple and transient hyper- and sub-aging.

The results link the hyper-aging exponent
$\delta'$ to the exponent characterizing the equilibrium
relaxation time.
Hence, they link a non-equilibrium dynamical exponent of the system
to a non-trivial equilibrium exponent, and through this to the
equilibrium static structure of the system.
Sub-aging on the other hand, emerges only as an effective cross-over,
\textit{i.e.}, as a finite-$t_w$ deviation from the mathematically rigorous
simple-aging asymptote. 

Interrupted hyper-aging versus sub-aging emerges as a clear indicator of the separation
between ideal-glass like dynamics, and the dynamics that arises from the
avoidance of the ideal glass transition. It could in principle be
used to determine more precisely the position of the ideal glass
transition.

This separation leads us to speculate that models
with a non-avoided \gls{MCT}-like glass transition
might show clear hyper-aging asymptotes.
High-dimensional systems of hard spheres, approaching the expected
mean-field-like behavior in $d=\infty$ \cite{Charbonneau.2017,Agoritsas.2019a},
might be suitable candidates.
On the other hand, in the context of spin glasses with \gls{MCT} transitions, \textit{e.g.},
the spherical $p$-spin model, numerical solutions so far favor
sub- and normal aging \cite{Cugliandolo.1993,Kim.2001,andreanov2006crossover}.
But the analytical determination of the scaling laws is still a critical open
issue \cite{Kim.2001,andreanov2006crossover}.
Hyper-aging in a trapped phase has been discussed very recently in
the context of decision-making models that incorporate reinforcement
by memory effects \cite{Moran.2020}.
Our Eq.~\eqref{eq:epstw} predicts weak ergodicity breaking and aging
that gets stuck at the \gls{MCT}-cricial point;
it will be interesting to explore
the connection to the
strong ergodicity breaking discussed in spin glasses \cite{Bernaschi.2020}
and the loss of ultrametricity connected with the hyper-aging asymptote
in suitably enhanced models.

\begin{acknowledgments}
We thank L.~Berthier, M.~Fuchs, and G.~Szamel for their valuable comments,
and A.~Meyer and M.~Medina-Noyola for continued support.
Part of this work has benefited from discussions at the
CECAM Flagship Workshop ``Memory Effects in Dynamical Processes''
of the Erwin-Schr\"odinger Institut (ESI) in Vienna.
Th.V.\ also thanks the Glass \& Time group at Roskilde University and
specifically Jeppe Dyre for their kind hospitality during a research visit
where this manuscript was finalized.
\end{acknowledgments}

\bibliographystyle{apsrev4-2}
\bibliography{references}
\end{document}